\documentclass[11pt,a4paper]{article}                           

\usepackage{a4}          

\usepackage[UKenglish]{babel}
\usepackage{amsmath}
\usepackage{amssymb}
\usepackage[usenames]{color} 
\usepackage{multirow}
\usepackage{graphicx}
\usepackage{epstopdf}
\usepackage{array}
\usepackage{hhline}
\usepackage{microtype}
\usepackage[bf]{caption}
\usepackage{color}
\usepackage[usenames,dvipsnames]{xcolor}
 
\usepackage{ifpdf}
\ifpdf
  \usepackage[pdftex,
    pdftitle={Vector-Like Exotics in F-Theory and 750 GeV Diphotons},
    pdfauthor={},
    pdfsubject={F-theory compactifications},
    bookmarksopen, bookmarksnumbered, bookmarksopenlevel=2]{hyperref}
\fi
\def\hybrid{\topmargin -20pt    \oddsidemargin 0pt
        \headheight 0pt \headsep 0pt
        \textwidth 6.25in       
        \textheight 9 in       
        \marginparwidth .875in
        \parskip 5pt plus 1pt 
          \jot = 1.5ex
   }
\hybrid
\numberwithin{equation}{section}
\numberwithin{table}{section}

\newcolumntype{C}[1]{>{\centering\let\newline\\\arraybackslash\hspace{0pt}}m{#1}}

\newcommand{\beq}{\begin{equation}}  \newcommand{\eeq}{\end{equation}}
\newcommand{\bal}{\begin{aligned}}   \newcommand{\eal}{\end{aligned}}
\newcommand{\bea}{\begin{eqnarray}}  \newcommand{\eea}{\end{eqnarray}}

\newcommand{\nn}{\nonumber}

\usepackage{cancel}

\newcommand{\be}{\begin{equation}}
\newcommand{\ee}{\end{equation}}

\begin{document}

\baselineskip=14pt
\parskip 5pt plus 1pt

\vspace*{-1.5cm}
\begin{flushright}    
  {\small
  }
\end{flushright}

\vspace{2cm}
\begin{center}        
  {\LARGE  Vector-Like Exotics in F-Theory and 750 GeV Diphotons}
\end{center}

\vspace{0.75cm}
\begin{center}        
Eran Palti
\end{center}

\vspace{0.15cm}
\begin{center}        
  {Institut f¨ur Theoretische Physik, Ruprecht-Karls-Universit¨at,
Philosophenweg 19, 69120, Heidelberg, Germany.}
             \\[0.15cm]
 
\end{center}

\vspace{2cm}


\begin{abstract}
\noindent
The recent excess in diphoton events around 750 GeV seen by the ATLAS and CMS experiments could be hinting at the existence of new vector-like charged matter around the TeV scale which couples to a singlet. Such a spectrum of exotics arises inevitably in certain classes of F-theory GUTs with hypercharge flux when the GUT symmetry is extended by a U(1) symmetry under which the Higgs fields of the MSSM are not vector-like. The exotics are not vector-like under the U(1) symmetry and therefore their mass is naturally related to its breaking scale. Previously this scale was taken to be close to the GUT scale which led to tension with proton decay, the $\mu$-term magnitude, and too large R-parity violation. The 750 GeV excess provides new motivation for considering breaking the U(1) around the TeV scale, which additionally alleviates the previous problems. We study the possible TeV-scale spectrum in such an SU(5) GUT scenario and show that it is constrained and predictive. Gauge coupling unification can be retained at the accuracy of the MSSM at one loop even though typically the spectrum does not form complete GUT representations. For example the exotics cannot form a complete 10 multiplet but nonetheless happen to behave as one in the beta functions. We present an initial analysis of the diphoton production rates for the exotics spectra and find them compatible with data. 
\end{abstract}

\thispagestyle{empty}
\clearpage

\setcounter{page}{1}


\newpage

\tableofcontents

\section{Introduction}

Recently the ATLAS and CMS experiments reported a possible excess in the diphoton spectrum near an invariant mass of 750 GeV \cite{atlas,cms}. An economic explanation for such an excess is a new scalar field with a mass around 750 GeV that decays to two photons. The scalar, S, should couple to gluons for production and photons for decay, and a particularly simple way to do so is through the introduction of some new vector-like matter which carries colour and electric charge and couples to the singlet. The singlet can then be produced and decay through loops involving this new vector-like matter. Of the large number of scenarios that have been proposed in the literature to explain the observed excess, see for example \cite{Franceschini:2015kwy,Knapen:2015dap}, a significant fraction are in this class. 

In this note we are interested in studying a possible ultraviolet origin for new vector-like charged matter near the TeV scale which couples to a singlet.\footnote{See \cite{Heckman:2015kqk,Cvetic:2015vit,Anchordoqui:2015jxc,Ibanez:2015uok} for alternative ideas on a string theory origin.} We retain the frameworks of low energy supersymmetry and high scale gauge coupling unification. Within this we are motivated by the observation that precisely such a spectrum of fields extending that of the Minimal Supersymmetric Standard Model (MSSM) arises inevitably when constructing a certain class of Grand Unified Theories (GUTs) in F-theory. F-theory is a very natural framework for implementing GUTs in string theory. One of the most interesting aspects of F-theory GUTs is that the breaking of the GUT symmetry group to that of the Standard Model is most naturally implemented through the introduction of hypercharge flux. In order for the hypercharge gauge field to stay massless this flux must be so-called globally trivial in F-theory \cite{Donagi:2008ca,Beasley:2008kw}. This restriction has been shown to impose quite strong constraints on the possible spectrum of particles whenever there is also an additional $U(1)$ symmetry present \cite{Marsano:2009wr,Dudas:2009hu,Dudas:2010zb,Marsano:2010sq,Palti:2012dd}. The restrictions on the spectrum can be argued for by an analysis of anomalies and are therefore quite robust. One particular implication, emphasised in \cite{Dudas:2010zb}, is that if one attempts to construct the spectrum of the MSSM where the Higgs up and Higgs down have different charges under the additional $U(1)$ then the MSSM spectrum is inevitably extended by additional massless vector-like charged matter and additional singlets which couple to this matter. This prediction of new vector-like and singlet matter from F-theory has so far been considered a problem, and the natural solutions implemented in the literature to it is to give the new singlets a GUT-scale vev thereby making the vector-like matter which couples to them very massive. Doing this however leads to tension with proton decay constraints, and often also with the magnitude of the $\mu$-term and matter parity violation, and so it is a simple though not completely clean solution \cite{Dudas:2010zb,Dolan:2011aq}. If the singlet vev, and therefore vector-like matter mass, is kept near the TeV scale the aforementioned problems are alleviated. While this forms a motivation for studying the possibility of keeping the exotics near the TeV scale in itself, the hints of an LHC excess that can be possibly explained by these exotics make exploring this scenario even more worthwhile. 

The presence of an additional $U(1)$ symmetry beyond the SM is also motivated by the $\mu$-problem of the MSSM. If the up and down Higgs fields have different charges under the symmetry then the superpotential quadratic term is forbidden and is replaced by a cubic term involving a $U(1)$ charged singlet, as in the NMSSM. A TeV-scale $\mu$-term can then be explained by a TeV-scale singlet vev. Therefore the breaking of the $U(1)$ symmetry occurs naturally at the TeV-scale in this scenario and since the new vector-like fields arise from hypercharge flux only due to the presence of precisely this $U(1)$ their mass can be naturally tied to the $\mu$-term scale in this way. It is important to note that the $U(1)$ itself can be, but need not be, massless. It depends on if it obtains a Stuckelberg mass in string theory which is always so if it is anomalous but can be so, depending on the model, even if it is not. Even if massive though it remains as a perturbative effective global symmetry.

The note is structured as follows. In section \ref{sec:ftrev} we review the form of theories coming from F-theory GUT constructions and in particular the restrictions which arise from hypercharge flux GUT breaking on the matter spectrum.  In section \ref{sec:minfth} we present the minimal implementation of a $U(1)$-extended $SU(5)$ GUT in F-theory where the Higgs up and down states have different $U(1)$ charges. We then study the spectrum of vector-like fields beyond the Minimal Supersymmetric Standard Model (MSSM) that arise in this construction as a results of hypercharge flux. In section \ref{genggc} we present a general analysis of the possible ways of obtaining a vector-like exotics spectrum that maintains gauge coupling unification and determine the possible charges under the U(1) that the spectrum then has. In section \ref{sec:extmod} we present an extended set of models which realise further phenomenological features. In section \ref{sec:prodrates} we present an initial analysis of diphoton prodcution rates for the exotics spectra. We summarise our results in section \ref{sec:summary}. 

\section{F-Theory spectra and hypercharge flux}
\label{sec:ftrev}

In this section we present a quick review of the matter spectrum of F-theory GUT models with hypercharge flux breaking. See \cite{Maharana:2012tu} for a more detailed review. We consider $SU(5)$ GUT constructions with additional (possibly massive) $U(1)$ symmetries. The GUT gauge symmetry is localised on a 7-brane wrapping a surface in the Calabi-Yau four-fold of F-theory. Matter is in the fundamental ${\bf 5}$ and anti-symmetric ${\bf 10}$ representations of the GUT group and in the absence of flux is vector-like. The possible different charges of the GUT matter under the additional $U(1)$ symmetries are model dependent. Chirality is generated in the spectrum after the introduction of $G$-flux along one of the $U(1)$ generators and the net number of generations for a given representation $R$ takes the form $\chi_R = Q_R\tilde{M}_R$ where $Q_R$ is the charge of the representation under the $U(1)$ symmetry and $\tilde{M}_R$ is a rational number such that the chirality is integer and corresponds to the geometric restriction of the flux to the locus where the matter is localised. Typically these would be curves inside the surface wrapped by the GUT 7-brane and we will therefore refer to these loci as matter curves and denote them by ${\cal C}_R^i$ where $R$ is the GUT representation and the index $i$ runs over the different U(1) charges present in the spectrum.

Once we introduce hypercharge flux the $SU(5)$ symmetry is broken to that of the Standard Model. The hypercharge flux can also modify the chirality of the matter spectrum in a certain restricted sense. If the matter loci have components which are so-called globally trivial then the chirality formula changes in a GUT non-universal way. The resulting chirality for a given GUT representation with universal $U(1)$ charge is
\bea
\# \left(3,1\right)_{-\frac13}-\# \left(\overline{3},1\right)_{\frac13} &=& M_{5} \;, \nn \\
\# \left(1,2\right)_{\frac12}- \# \left(1,2\right)_{-\frac12}  &=& M_{5} + N _{5}\;, \label{chi5sm}
\eea
for the 5-matter curves and
\bea
\# \left(3,2\right)_{\frac16}-\# \left(\overline{3},2\right)_{-\frac16}  &=& M_{10} \;, \nn \\
\# \left(\overline{3},1\right)_{-\frac23}-\# \left(3,1\right)_{\frac23}  &=& M_{10} - N_{10} \;, \nn \\
\# \left(1,1\right)_{+1}-\# \left(1,1\right)_{-1} &=& M_{10} + N_{10}\;, \label{chi10sm}
\eea
for the 10-matter curves. Here $N_R$ is an integer associated to the hypercharge flux restriction and $M_R$ is an integer associated to the U(1) flux restriction. The interesting restrictions on the spectrum amount to relations between the values of the $N_R$ integers for different representations. They were first noted as geometric constraints in \cite{Marsano:2009wr,Dudas:2009hu,Dudas:2010zb} and later interpreted as related to anomalies \cite{Marsano:2010sq,Palti:2012dd}. When considering all the matter representations in the spectrum the restrictions on the flux values read
\bea
& &\sum_{{\cal C}_{10}^i} Q_{10}^i N_{10}^i + \sum_{{\cal C}_{5}^j} Q_{5}^j N_{5}^j= 0 \;, \label{anomNmix}  \\
& &\sum_{{\cal C}_{10}^i} N_{10}^i = \sum_{{\cal C}_{5}^j} N_{5}^j = 0 \;, \label{nochiN}  \\
& &3 \sum_{{\cal C}_{10}^i} \left(Q_{10}^i\right)^2 N_{10}^i + \sum_{{\cal C}_{5}^j} \left(Q_{5}^j\right)^2 N_{5}^j = 0 \;. \label{anomN}
\eea
The interpretation in terms of anomalies is that the $U(1)$ symmetry can be anomalous but only in a GUT universal way which means that the introduction of the hypercharge flux should not modify the anomalies \cite{Marsano:2010sq,Palti:2012dd}. This is because the global triviality of the hypercharge flux means it can not participate in the Green-Schwarz (GS) anomaly cancellation mechanism of string theory.\footnote{For an F-theory study of the GS mechanism see \cite{Cvetic:2012xn}.} In essence this particular restriction on the spectrum, between full anomaly cancellation as usually considered in field theory and no restriction from Abelian anomalies as usually expected in string theory (due to GS closed-string anomaly cancellation), leads to the signature charges of the matter which we explore in this note.

In \cite{Dudas:2010zb} it was shown that (\ref{anomNmix}) implies that if the Higgs up and down states of the MSSM spectrum have different charges under the $U(1)$ symmetry then the MSSM spectrum must be extended by additional states which are vector-like under the SM gauge symmetries but not under the $U(1)$. This prediction of necessary additional vector-like charged fields beyond the MSSM spectrum, and singlets which couple to them, forms one of the key motivations for this note in light of the possible 750 GeV excess. 

The previous discussion amounts to general statements about possible F-theory models, let us summarise the status of actual explicit examples manifesting these properties. There are two important aspects for different F-theory models relevant to the analysis in this note: the U(1) charges of the spectrum and the restriction of the hypercharge flux to the matter loci. Before considering hypercharge flux there are two types of U(1) charges we will study, the first type of models are those for which we have a complete well-understood global F-theory geometry, and in particular know the associated smooth resolved geometry in which the charges can be calculated, we will call these global smooth models. The second type of models we will consider are ones presented in \cite{Baume:2015wia}. These models are natural from an F-theory perspective though we do not know a smooth resolved geometry associated to them. Nonetheless they can be reasonably expected to arise from F-theory in the sense that in \cite{Baume:2015wia} a Higgsing chain to reach them from certain so-called factorised Tate models was presented, and the singular elliptic fibrations associated to these factorised Tate models were presented in \cite{Mayrhofer:2012zy}. They are also consistent with, and form a subset of, the general analysis of possible U(1) charges in global F-theory geometries \cite{Kuntzler:2014ila,Lawrie:2014uya,Lawrie:2015hia}. 

With regards to hypercharge flux, the constraints  (\ref{anomNmix})- (\ref{anomN}) which we will consider amount to a distribution of hypercharge flux over different U(1) charged matter loci. As yet we only have a local understanding of this possibility but a global implementation is still missing. Note that the difficult part is implementing a net hypercharge flux restriction to a matter state with a U(1) charge, while a global implementation of hypercharge flux which has no net restriction to a U(1) charged state is much simpler to implement, see for example \cite{Braun:2014pva}. We therefore assume in this work that the lack of global examples with net hypercharge flux restriction is not due to some unknown obstruction but only due to technical difficulty. From a local approach it appears that the constraints  (\ref{anomNmix})- (\ref{anomN}) are stronger than any geometric constraints known, ie. local geometries which realise the maximal range of hypercharge fluxes compatible with the constraints can typically be constructed, see for example \cite{Dolan:2011iu}. 

Global examples with net hypercharge restriction to matter loci were constructed in the weakly-coupled type IIB limit of F-theory in \cite{Mayrhofer:2013ara}. However the uplift of these constructions to F-theory remains poorly understood. Importantly it was shown that in the constructions of \cite{Mayrhofer:2013ara} there exist configurations where the hypercharge flux can modify the anomalies, and therefore avoid the constraints (\ref{anomNmix})- (\ref{anomN}), but only in a very restrictive way involving so-called geometrically massive U(1)s \cite{Grimm:2011tb}. Therefore the anomalies with respect to the geometrically massless U(1)s remain unmodified and the constraints (\ref{anomNmix})-(\ref{anomN}) indeed are observed in the IIB constructions. The {\it class} of F-theory models that we will study therefore, as referred to in the abstract, are ones where hypercharge flux can be distributed with net restriction to matter curves and where the $U(1)$s are not geometrically massive, or at least do not have their Green-Schwarz anomaly cancellation mechanism modified by hypercharge flux.\footnote{We note that the uplift of geometrically massive U(1)s to F-theory is not fully understood \cite{Grimm:2010ez,Grimm:2011tb,Anderson:2014yva,Mayrhofer:2014laa} but an analysis of similarly massive U(1)s in \cite{Garcia-Etxebarria:2014qua,Mayrhofer:2014laa,Mayrhofer:2014haa,Cvetic:2015moa,Klevers:2014bqa,Martucci:2015oaa} leads one to expect that they do not lead to full U(1) selection rules but at best to some discrete symmetry remnant. Therefore they are not as well motivated as protection mechanisms, for example in protecting the $\mu$ term.}

An important role will be played by fields that are singlets under the SM gauge group but charged under the extending U(1) groups. These arise generically in F-theory constructions with U(1) symmetries, with the general rule that a singlet is present whenever it can form a U(1) neutral Yukawa coupling with a pair of 5-matter representations of type $1 5 \overline{5}$. Turning on flux along the U(1) can induce chirality for these fields which may be important as a mechanism for keeping them massless, though this is a model dependent statement. There are no universal constraints from anomalies on the spectrum of singlet fields because the U(1) can itself be anomalous. We will therefore in general leave their chirality as free parameters. 

So far we have discussed exotic fields extending the spectrum of the MSSM that are induced by hypercharge flux restriction to matter loci in F-theory. There is a qualitatively different set of possible exotics that are induced by hypercharge flux which are not related to the matter loci but come from so-called bulk modes. These are interesting because they arise inevitably if the GUT group is broken by fluxes, including hypercharge flux, directly to the Standard Model gauge group for GUT groups larger than SU(5) such as SO(10) and $E_6$ \cite{Beasley:2008kw}. However such bulk exotics are always vector-like under all the symmetries including the U(1). This means that while they might be massless in certain vacua there is no symmetry to protect their mass. This may not be a serious issue as they could happen to obtain masses at the TeV scale only, but this is a model-dependent statement. It is worth noting however, in light of the 750 GeV excess, that it is possible to induce bulk exotics coming from hypercharge flux breaking of SU(5) in the representation $\left(3,2\right)_{-\frac56}$ which is appealing due to its large hypercharge (and electric charge) quantum number which enhances the diphoton production rate. However even if they were assumed to only pick up a mass at the TeV scale there would still be two crucial obstructions to considering them as candidates: The first is that they are uncharged under the U(1) which means that they must couple to some completely neutral singlet. The second is that there is no way to preserve perturbative gauge coupling unification with such a vector pair. Nonetheless we mention them here as a possibility not to be completely discounted.

\section{The minimal F-theory $U(1)$ model}
\label{sec:minfth}

In this section we consider what is in some sense the minimal F-theory model with a $U(1)$ such that the Higgs up and down have different charges. This is the so called $U(1)_{PQ}$ model which was first studied from a local perspective in \cite{Marsano:2009wr} and whose global realisation was presented in \cite{Mayrhofer:2012zy}. The spectrum of states in the model is shown in table \ref{u1emb2}. As stated previously, the SU(5) model is completely well-understood as a global F-theory construction, however the net restriction of the hypercharge flux to the matter curves is only understood from a local perspective and while there are no known obstructions as yet there are no explicit global examples of this. With this caveat in mind we proceed and the possible hypercharge restrictions to the matter curves that are compatible with the set of restrictions (\ref{anomNmix}-\ref{anomN}), as calculated in \cite{Palti:2012dd}, is also shown in table \ref{u1emb2}. 
\begin{table}
\centering
\begin{tabular}{|c|c|c|c|}
\hline
Field & $U(1)$ Charge & $U(1)$ Flux & Hypercharge Flux \\
\hline
$10_E$ & -2 & $M_{10_E}$ & N \\
\hline
$10_M$ & 3 & $M_{10_M}$ & -N \\
\hline
$5_{H_d}$ & 4 & $M_{5_{H_d}}$ & N \\
\hline
$5_{M}$ & -1 & $M_{5_{M}}$ & -N \\
\hline
$5_{H_u}$ & -6 & $M_{5_{H_u}}$ & 0 \\
\hline
\end{tabular}
\caption{Table showing one-parameter family of hypercharge flux configurations compatible with anomalies for the minimal F-theory $U(1)$ model.}
\label{u1emb2}
\end{table}
The minimal model which exhibits doublet-triplet splitting is obtained by taking the flux choices in table \ref{u1minspect} where the resulting spectrum is shown.
\begin{table}
\centering
\begin{tabular}{|c|c|c|c|c|c|}
\hline
GUT Field & $U(1)$ Charge& $U(1)$ Flux & Hyper Flux & MSSM & Exotics \\
\hline
$10_E$ & -2 & -1 & -1 & & $\left(\bar{3},2\right)_{-\frac16} + 2 \times\left(1,1\right)_{-1}$\\
\hline
$10_M$ & 3 & 4 & 1 &$3 \times 10$ & $\left(3,2\right)_{\frac16} + 2 \times\left(1,1\right)_1$\\
\hline
$5_{H_d}$ & 4 & 0& -1&$\left(1,2\right)_{-\frac12}$ & \\
\hline
$5_{M}$ & -1 & -4 & 1 & $3\times \overline{5}$ & $\left(\bar{3},1\right)_{\frac13}$ \\
\hline
$5_{H_u}$ & -6 & 1 & 0&$\left(1,2\right)_{\frac12}$ & $\left(3,1\right)_{-\frac13}$\\
\hline
$S_1$ & 5 &  & & &$\left(1,1\right)_{0}$\\
\hline
$S_2$ & 10 &  & & & $\left(1,1\right)_{0}$\\
\hline
\end{tabular}
\caption{Table showing the massless spectrum for the minimal F-theory $U(1)$ model.}
\label{u1minspect}
\end{table}
The spectrum is that of the MSSM with additional vector-like fields
\bea
& &E_1 = \left(3,2\right)^{3}_{\frac16} \;,\;\; E_{2,1} = \left(1,1\right)^3_1 \;,\;\; E_{2,2} = \left(1,1\right)^3_1\;,\;\; E_3 = \left(3,1\right)^{-6}_{-\frac13} \;, \nonumber \\
& & \overline{E}_1 = \left(\overline{3},2\right)^{2}_{-\frac16} \;,\;\; \overline{E}_{2,1} = \left(1,1\right)^2_{-1} \;,\;\; \overline{E}_{2,2} = \left(1,1\right)^2_{-1}\;,\;\; \overline{E}_3 = \left(\bar{3},1\right)^{1}_{\frac13} \;. \label{minfexo}
\eea
The subscripts in (\ref{minfexo}) denote the hypercharges while the superscripts denote the U(1) charges. There are also SM singlet fields charged under the U(1) as shown in table \ref{u1minspect}, as well as their conjugate oppositely charged partners $\overline{S}_1$ and $\overline{S}_2$
\be
S_1 = \left(1,1\right)^{5}_{0} \;,\;\; \overline{S}_1 = \left(1,1\right)^{-5}_{0} \;,\;\; S_2 = \left(1,1\right)^{10}_{0} \;,\;\; \overline{S}_2 = \left(1,1\right)^{-10}_{0} \;.
\ee
The relevant renormalisable superpotential terms are schematically  
\be
W \supset \overline{S}_1 \left(E_1 \overline{E}_1 + E_{2,1} \overline{E}_{2,1} + E_{2,2} \overline{E}_{2,2}\right) + S_2 5_{H_u} \bar{5}_{H_d} + S_1 \left(5_{H_u} \bar{5}_{M} + E_3 \overline{E}_3\right) + S_1^2 \overline{S}_2 +  \overline{S}_1^2 S_2 \;. \label{minfsup}
\ee
We have suppressed the coupling constants which appear for each of the terms in (\ref{minfsup}). At this point we would like to consider the possible vacuum expectation values for the scalar components of the singlets $S_i$ and $\overline{S}_i$. The potential for these fields is rather complicated to calculate and highly model dependent, it receives contributions from soft supersymmetry breaking masses, D-terms of the U(1)s, non-perturbative effects in string theory, and couplings between the singlets. We therefore at this stage take their vevs to be free parameters rather than attempting to fix them dynamically. We see from the superpotential interactions that all the exotic fields can obtain a mass set by the vev of the singlets $S_1$ and $\overline{S}_1$. Typically we might consider taking their vev very large, of order the GUT scale, to decouple the exotics. However we see that there are already some problems with doing so: we would then generate a very large matter-parity violating term $5_{H_u} \overline{5}_{M}$ which would induce, among other things, large neutrino masses, and also a very large mass for the $\mu$-term singlet $S_2$ which can obstruct it obtaining a TeV-scale vev. Further problems arise when considering the non-renormalisable couplings
\be
W \supset \overline{S}_1 \overline{5}_M \overline{5}_M 10_M + \overline{S}_2 \overline{5}_M 10_M 10_M 10_M \;.
\ee
These couplings lead to proton decay for large vevs of the singlets.\footnote{The dimension 5 proton decay operator also has a coupling to $\frac{\overline{S}_1}{S_1}$ once the exotics are integrated out which limits the ratio of the singlet vevs to not be too large.} This situation is fairly typical in F-theory constructions: it is difficult to satisfy phenomenological constraints while decoupling the exotics at a very high scale. See a review of this problem in \cite{Maharana:2012tu}. There is though a rather simple solution which is that the singlets only obtain a vev around the TeV scale. As a first approximation this is sufficient to satisfy the phenomenological constraints.\footnote{The R-parity violating operator is very constrained and would require a further suppression by some orders of magnitude which might be possible through small couplings.} We do not consider the fine details of the phenomenology at this point, especially since this is the minimal model and more sophisticated versions can be developed. Rather the main point of the model is to illustrate the issues that arise rather generically in these constructions and how they enforce a light mass on the exotics with respect to the GUT scale. The result of these considerations is that the exotic spectrum, and the singlets that couple to it, can be motivated quite generally to be sitting around the TeV scale, something which receives further motivation by the possible excess at 750 GeV observed at the LHC \cite{atlas,cms}.

One rather immediate objection from a theoretical perspective to keeping the exotic spectrum light is that it does not form complete GUT multiplets and therefore is expect to ruin gauge coupling unification. Given that the starting point of the F-theory constructions is the GUT paradigm this is a valid point. However, remarkably the spectrum of exotics behaves the same as a pair of complete 10 representations in the 1-loop beta functions
\be
10=\left[ (3,2)_{+\frac16} + (\bar{3},1)_{-\frac23} + (1,1)_{+1} \right] \
 \sim \ \left[ (3,2)_{+\frac16} + (\bar{3},1)_{+\frac13} + 2\times(1,1)_{+1} \right] \;, \label{specexot2}
\ee
and therefore gauge coupling unification is preserved to the same accuracy as the MSSM at one loop! It is not always the case that for any F-theory model the spectrum of exotics has this property of preserving unification. However, as this example shows, such combinations arise rather naturally and generically, as first noted in \cite{Dudas:2010zb}. Note that gauge coupling unification also constrains the vev of the singlets $S_1$ and $\overline{S}_1$ to be similar in magnitude.

It is worth noting that the feature of this model that the mass of the exotics is set by the vev of a singlet $S_1$ and its conjugate state $\overline{S}_1$ has positive and negative aspects. On the positive side the D-term constrain motivates, though not necessarily implies, that their vevs are equal which helps guarantee gauge coupling unification. This is because we can not protect the lightness of the singlets themselves in a natural way using chirality. It is possible in principle that they are chiral, ie. only say $S_1$ is massless, but then it may be difficult to justify a TeV-scale vev for the massive singlet. Perhaps a more natural possibility is that they are vector-like but happen to be massless, this happens often in string theory in certain supersymmetric vacua, and then one expects such massless modes to be generally lifted by other effects such as supersymmetry breaking. Such a mechanism could induce TeV-scale masses for them though overall is less appealing than protecting their mass through chirality. We note that while this vector-like structure for the singlets that lift the exotics is a feature of the minimal model it is not universal, indeed we present example models without this feature in section \ref{sec:extmod}, and the general analysis of when this occurs can be deduced from the analysis of section \ref{genggc}.

The exotics spectrum is somewhat unusual when considered in a GUT context, and also it has a very particular coupling structure to the singlet fields due to the U(1) symmetry. These properties can serve as quite distinctive signatures of such a spectrum. We now proceed to analyse in general these particular structures of exotics in F-theory: how to obtain spectra that are consistent with gauge coupling unification and what are their signature charges under the U(1) symmetry.

\section{General constraints on the spectrum and gauge coupling unification}
\label{genggc}

In this section we analyse in generality the system of equations (\ref{chi5sm}-\ref{chi10sm}) and (\ref{anomNmix}-\ref{anomN}). In particular we are interested in the possibility of preserving gauge coupling unification. This means that the spectrum, which is automatically vector-like from (\ref{nochiN}), can either form complete GUT multiplets or can form combinations which act as complete multiplets in the beta functions. We split the analysis into the cases where there is some hypercharge restriction to 10 matter loci, and where there is no such restriction.  It is worth noting at this point that in addition to the spectrum which is induced inevitably by hypercharge flux which is vector-like with respect to the Standard Model gauge groups but not with respect the the additional $U(1)$, there may also be an arbitrary number of multiplets which are vector-like under all the symmetries of the theory. We do not discuss such states as their mass is not protected by any symmetry and because their presence is highly model-dependent.

In this section we classify the possible ways to induce the spectrum of exotics and the constraints on their charges. This must then be combined with an additional MSSM spectrum and implemented on appropriately charges matter curves to generate the necessary Yukawa couplings. In terms of the spectrum we account for the vector-like Higgs doublets in this section, since their presence is tied to hypercharge flux, but not for the 3 full generations.\footnote{This assumes that the MSSM multiplets come from complete GUT multiplets, while more general possibilities of splitting them up were studied for example in \cite{Krippendorf:2014xba,Krippendorf:2015kta}.} The results of this section can then form a starting point for more complete model building as studied in sections \ref{sec:minfth} and \ref{sec:extmod}.

\subsection{Hypercharge restriction to 10 matter}

From (\ref{chi10sm}) we see that it is not possible for the exotics spectrum to form a single vector-like pair of complete 10-multiplets because the chiral difference between the states $(\bar{3},1)_{-2/3}$ and $\left(1,1\right)_1$ is minimally 2. This means that the minimal spectrum which forms complete multiplets is two complete vector-like pairs, which is incompatible with perturbative unification of the gauge couplings if they have a mass around the TeV scale. 

If we have a hypercharge restriction of $N=1$ in (\ref{chi10sm}) and we take $M=1$ we find the spectrum $(3,2)_{\frac16} + 2\times (1,1)_1$. If we also have a restriction to a 5 matter states with $M=-1$ and $N=1$ we get a state $(\bar{3},1)_{\frac13}$. Now as discussed in section \ref{sec:minfth} this combination of states, which as we have seen arises quite naturally, acts as a complete 10 representation in the 1-loop beta function. 
Generally, the combination of states (\ref{specexot2}) can in principle arise by hypercharge restriction to two 10 matter curves and zero hypercharge restriction to any 5 matter curves. In this case the states $(\bar{3},1)_{\frac13}$ would be the GUT partners of the Higgs doublets. However from the constraint (\ref{anomNmix}) we see that this is not possible, since a sole restriction to two 10 matter curves would imply that they must have equal charge under the $U(1)$. It is possible to realise just the spectrum (\ref{specexot2}) using only two 10 matter curves and two 5 matter curves. However this is very constrained and therefore we also consider the case of two 10 matter curves and three 5 matter curves. 

We now wish to classify the possible fluxes and U(1) charges that can lead to the spectrum of exotics (\ref{specexot2}) and the MSSM Higgses. This can be used as a tool for model building, but also serves as a predictive framework for the spectrum which is experimentally accessible. The first step is classifying solutions to the constraints (\ref{anomNmix}-\ref{anomN}) for hypercharge flux distributed over two 10 mater curves and three 5 matter curves. In principle there is an infinite number of solutions to the constraints but only if we allow the U(1) charges of the states to vary over an infinite range. This is not likely to be possible in F-theory. As yet there is no complete classification of the possible U(1) charges of states in F-theory. In \cite{Lawrie:2015hia} a classification was derived of possible charges in certain smooth cases, though it was not determined if all the charges can appear simultaneously in a single model. As yet there are no examples known in F-theory where the SU(5) charged spectrum is larger than that which comes from a single adjoint representation of $E_8$. This adjoint has five 10-matter states and ten 5-matter states which fill out a lattice of charges. A rank one top quark Yukawa coupling, as well as breaking the four extra U(1)s to the single U(1) case studied here will require this charge spectrum to be reduced. In our analysis we therefore make the assumption that in the model there are at most 5 10-curves and 8 5-curves which fully occupy a lattice of charges, ie.  in appropriate units the charges for each type of matter state fill out the appropriate range of natural numbers. With the relatively mild assumption discussed above we are able to classify all the possibilities. The discussion of the methodology used is presented in appendix \ref{sec:app}, while the results are presented in table \ref{u1char10}.

Once we have a spectrum of exotics which acts as a vector-like pair of 10 representations we can only add one further vector-like pair of the 5 representation while maintaining perturbative gauge coupling unification. This possibility is also classified in table \ref{u1char10}.
The final combination of states that preserves gauge coupling unification acts in the same way in the 1-loop beta functions as a pair of 10 and 5 multiplets
\be
10+5 
 \sim  \left[ \left(3,2\right)_{+\frac16} + 2 \times \left(3,1\right)_{+\frac23} + \left(1,2\right)_{+\frac12}\right] \;. \label{specexot1}
\ee
It is worth noting that with respect to the 750 GeV excess such a combination of states may be favourable over the complete multiples due to the higher hypercharge values. 
A vector-pair of these representations, and the MSSM Higgs vector-pair, can be realised over two 10 curves and two 5 curves. In table \ref{u1char10} we present this configuration as well as the possible realisations of this spectrum over two 10 curves and three 5 curves.
\def\arraystretch{1.0}
{\tiny
\begin{table}
\small
\centering
\begin{tabular}{|C{6.0cm}|C{4cm}|C{5.0cm}|}
\hline
$U(1)$-Fluxes $\left\{M_{10}^1,M_{10}^2,M_5^1,M_5^2,M_5^3\right\}$ & Hypercharge $\left\{N_{10}^1,N_{10}^2,N_5^1,N_5^2,N_5^3\right\}$ & Charge Constraints  $\left\{a,b,c,d\right\}$\\
\hline
\hline
\multicolumn{3}{|c|}{Exotics: $\left[ (3,2)_{+\frac16} + (\bar{3},1)_{+\frac13} + 2\times(1,1)_{+1} \right]$+ Conjugate $\sim 10+\overline{10}$} \\
\hline
$\left\{1,-1,1,-1,0\right\}$ & $\left\{1,-1,-2,2,0\right\}$ & $\left\{4,7,5,-\right\}$\\
\hline
$\left\{1,-1,0,-1,1\right\}$,$\left\{1,-1,1,0,-1\right\}$ & $\left\{1,-1,-1,1,0\right\}$ & $\left\{1,2,1,-\right\}$\\
\hline
$\left\{-1,1,-1,1,0\right\}$,$\left\{-1,1,0,1,-1\right\}$ & $\left\{-1,1,1,-2,1\right\}$ & $\left\{1,1,-1,-2\right\}$,$\left\{1,-2,-5,-7\right\}$,\newline$\left\{1,-7,-11,-14\right\}$\\
\hline
$\left\{1,-1,-1,1,0\right\}$,$\left\{1,-1,0,1,-1\right\}$ & $\left\{1,-1,1,-2,1\right\}$ & $\left\{1,5,4,2\right\}$,$\left\{1,10,8,5\right\}$,\newline$\left\{1,17,14,10\right\}$\\
\hline
\hline
\multicolumn{3}{|c|}{Exotics: $\left[ (3,2)_{+\frac16} + (\bar{3},1)_{+\frac13} + 2\times(1,1)_{+1} +5\right]$+ Conjugate $\sim 10+\overline{10}+5+\overline{5}$} \\
\hline
$\left\{1,-1,2,-2,0\right\}$ & $\left\{1,-1,-4,4,0\right\}$ & $\left\{8,13,11,-\right\}$\\
\hline
$\left\{1,-1,0,-2,2\right\}$,$\left\{1,-1,2,0,-2\right\}$ & $\left\{1,-1,-2,2,0\right\}$ & $\left\{4,7,5,-\right\}$\\
\hline
$\left\{1,-1,1,-2,1\right\}$,$\left\{1,-1,2,-1,-1\right\}$ & $\left\{1,-1,-3,3,0\right\}$ & $\left\{3,5,4,-\right\}$\\
\hline
$\left\{-1,1,-2,1,1\right\}$,$\left\{-1,1,0,2,-2\right\}$ & $\left\{-1,1,2,-3,1\right\}$ & $\left\{1,1,0,-1\right\}$,$\left\{3,5,3,2\right\}$,\newline$\left\{1,-5,-7,-10\right\}$\\
\hline
$\left\{-1,1,-2,1,1\right\}$,$\left\{-1,1,1,1,-2\right\}$ & $\left\{-1,1,1,-2,1\right\}$ & $\left\{1,1,-1,-2\right\}$,$\left\{1,-2,-5,-7\right\}$,\newline$\left\{1,-7,-11,-14\right\}$\\
\hline
$\left\{-1,1,-2,2,0\right\}$,$\left\{-1,1,-1,2,-1\right\}$ & $\left\{-1,1,3,-4,1\right\}$ & $\left\{2,3,2,1\right\}$,$\left\{1,-1,-2,-4\right\}$,\newline$\left\{3,3,1,-2\right\}$,$\left\{1,-15,-17,-22\right\}$\\
\hline
$\left\{-1,1,-2,2,0\right\}$,$\left\{-1,1,1,1,-2\right\}$ & $\left\{-1,1,1,-3,2\right\}$ & $\left\{1,-1,-4,-5\right\}$,$\left\{3,5,0,-1\right\}$,\newline$\left\{1,-10,-15,-17\right\}$\\
 \hline
$\left\{-1,1,-2,2,0\right\}$,$\left\{-1,1,-1,2,-1\right\}$,\newline$\left\{-1,1,0,2,-2\right\}$ & $\left\{-1,1,2,-4,2\right\}$ & $\left\{4,5,1,-1\right\}$\\
\hline
$\left\{-1,1,-1,2,-1\right\}$,$\left\{-1,1,0,2,-2\right\}$ & $\left\{-1,1,1,-4,3\right\}$ & $\left\{1,-4,-8,-9\right\}$,$\left\{2,1,-4,-5\right\}$,\newline$\left\{3,4,-2,-3\right\}$\\
\hline
$\left\{1,-1,-2,1,1\right\}$,$\left\{1,-1,1,1,-2\right\}$ & $\left\{1,-1,1,-2,1\right\}$ & $\left\{1,5,4,2\right\}$,$\left\{1,10,8,5\right\}$,\newline$\left\{1,17,14,10\right\}$\\
\hline
$\left\{1,-1,-2,1,1\right\}$,$\left\{1,-1,0,2,-2\right\}$ & $\left\{1,-1,2,-3,1\right\}$ & $\left\{1,8,7,4\right\}$,$\left\{3,10,9,4\right\}$,\newline$\left\{1,20,18,13\right\}$\\
\hline
$\left\{1,-1,-2,2,0\right\}$,$\left\{1,-1,1,1,-2\right\}$ & $\left\{1,-1,1,-3,2\right\}$ & $\left\{1,4,3,2\right\}$,$\left\{3,7,6,4\right\}$,\newline$\left\{1,13,10,8\right\}$\\
\hline
$\left\{1,-1,-2,2,0\right\}$,$\left\{1,-1,-1,2,-1\right\}$ & $\left\{1,-1,3,-4,1\right\}$ & $\left\{1,12,11,7\right\}$,$\left\{2,11,10,5\right\}$,\newline$\left\{3,12,11,5\right\}$\\
\hline
$\left\{1,-1,-2,2,0\right\}$,$\left\{1,-1,-1,2,-1\right\}$,\newline$\left\{1,-1,0,2,-2\right\}$ & $\left\{1,-1,2,-4,2\right\}$ & $\left\{4,13,11,7\right\}$\\
\hline
$\left\{1,-1,-1,2,-1\right\}$,$\left\{1,-1,0,2,-2\right\}$ & $\left\{1,-1,1,-4,3\right\}$ & $\left\{2,5,4,3\right\}$,$\left\{1,7,5,4\right\}$,\newline$\left\{3,11,8,6\right\}$,$\left\{1,25,20,18\right\}$\\
\hline
$\left\{1,-1,-1,-1,2\right\}$,$\left\{1,-1,0,-2,2\right\}$\;,\newline$\left\{1,-1,1,1,-2\right\}$,$\left\{1,-1,2,0,-2\right\}$ & $\left\{1,-1,-1,1,0\right\}$ & $\left\{1,2,1,-\right\}$\\
\hline
\hline
\multicolumn{3}{|c|}{Exotics: $\left[ (3,2)_{+\frac16} + 2 \times  (3,1)_{+\frac23} + (1,2)_{+\frac12}\right]$+ Conjugate $\sim 10+\overline{10}+5+\overline{5}$} \\
\hline
$\left\{-1,1,0,0,0\right\}$ & $\left\{1,-1,-2,2,0\right\}$ & $\left\{4,7,5,-\right\}$\\
\hline
$\left\{-1,1,0,0,0\right\}$ & $\left\{1,-1,1,-2,1\right\}$ & $\left\{1,5,4,2\right\}$,$\left\{1,10,8,5\right\}$,\newline$\left\{1,17,14,10\right\}$ \\
\hline
\end{tabular}
\caption{Table showing flux configurations which lead to exotic spectra consistent with gauge coupling unification and the constraints on the U(1) charges of the states. The notation for the last column is such that the charge relations are $Q_{10}^1 = a r +Q_{10}^2\;,\;Q_5^1 = b r +3Q_{10}^2\;,\;Q_5^2 = c r+3Q_{10}^2$, and if $N_5^3$ is non-vanishing $Q_5^3 = d r+3Q_{10}^2$. Where $r$ is an arbitrary rational number constrained such that the charges are integer.}
\label{u1char10}
\end{table}
\normalsize
\def\arraystretch{1.0}

\subsection{Hypercharge restriction to 5 matter}
\label{sec:hyp5comp}

In the previous section we studied the possibilities of the spectrum when there is some hypercharge restriction to the 10-matter curves. In this section we will consider the spectrum that can arise when the hypercharge restricts to only the 5-matter curves. In some sense this is the most minimal extension to the MSSM spectrum. The only possibility of preserving gauge coupling unification in this case is through complete 5 multiplets. Using the constraints (\ref{anomNmix}-\ref{anomN}) we can deduce that the minimal number of curves that the hypercharge flux must restrict to to get full 5 multiplets is 4. We can constrain the possible U(1) charges that the states can carry for this minimal implementation over 4 curves. First note that we can perform a universal shift in the U(1) charges of all the 5 matter states whilst leaving the constraints (\ref{anomNmix}-\ref{anomN}) unchanged. Similarly we can rescale all the charges universally. Therefore we are free, with full generality for the purpose of this analysis, to take the U(1) charges of the 5-matter states to be the natural numbers $\left\{0,1,2,...\right\}$. We can then scan over the possible hypercharge and U(1) flux assignments that can lead to an exotic spectrum of a maximum number of 4 complete 5 multiplets (note that we must also make sure of the presence of a vector-pair of doublets to form the Higgs fields of the MSSM). We present the results in table \ref{u1charcomp5}. 
\begin{table}
\centering
\begin{tabular}{|C{5.5cm}|C{5cm}|C{4cm}|}
\hline
Charges $\left\{Q_1,Q_2,Q_3,Q_4\right\}$ & $U(1)$-Fluxes $\left\{M_1,M_2,M_3,M_4\right\}$ & Hypercharge $\left\{N_1,N_2,N_3,N_4\right\}$ \\
\hline
\hline
\multicolumn{3}{|c|}{1 vector pair of 5 multiplets} \\
\hline
\hline
$\left\{0,1,3,4\right\}$ & $\left\{0,-1,1,0\right\}$,$\left\{0,0,1,-1\right\}$,\newline$\left\{1,-1,0,0\right\}$,$\left\{1,0,0,-1\right\}$ & $\left\{-1,2,-2,1\right\}$   \\
\hline
\hline
\multicolumn{3}{|c|}{2 vector pairs of 5 multiplets} \\
\hline
\hline
$\left\{0,1,5,6\right\}$ & $\left\{2,0,0,-2\right\}+...\; (9)$ & $\left\{-2,3,-3,2\right\}$   \\
\hline
$\left\{0,1,3,4\right\}$ & $\left\{2,0,0,-2\right\}+...\; (9)$ & $\left\{-1,2,-2,1\right\}$   \\
\hline
$\left\{0,1,2,3\right\}$ & $\left\{2,-1,0,-1\right\}+...\; (9)$ & $\left\{-1,3,-3,1\right\}$   \\
\hline
$\left\{0,3,5,8\right\}$ & $\left\{1,-1,1,-1\right\}+...\; (9)$ & $\left\{-1,4,-4,1\right\}$   \\
\hline
\hline
\multicolumn{3}{|c|}{3 vector pairs of 5 multiplets} \\
\hline
\hline
$\left\{0,1,7,8\right\}$ & $\left\{3,0,0,-3\right\}+...\; (16)$ & $\left\{-3,4,-4,3\right\}$   \\
\hline
$\left\{0,1,5,6\right\}$ & $\left\{3,0,0,-3\right\}+...\; (15)$ & $\left\{-2,3,-3,2\right\}$   \\
\hline
$\left\{0,1,3,4\right\}$ & $\left\{3,0,0,-3\right\}+...\; (22)$ & $\left\{-1,2,-2,1\right\}$   \\
\hline
$\left\{0,1,2,3\right\}$ & $\left\{3,-1,0,-2\right\}+...\; (14)$ & $\left\{-1,3,-3,1\right\}$   \\
\hline
$\left\{0,3,5,8\right\}$ & $\left\{3,-2,0,-1\right\}+...\; (12)$ & $\left\{-1,4,-4,1\right\}$   \\
\hline
$\left\{1,3,4,6\right\}$ & $\left\{-1,-2,3,0\right\}+...\; (8)$ & $\left\{-1,5,-5,1\right\}$   \\
\hline
\hline
\multicolumn{3}{|c|}{4 vector pairs of 5 multiplets} \\
\hline
\hline
$\left\{0,1,7,8\right\}$ & $\left\{4,0,0,-4\right\}+...\; (19)$ & $\left\{-3,4,-4,3\right\}$   \\
\hline
$\left\{0,1,4,5\right\}$ & $\left\{4,-1,0,-3\right\}+...\; (16)$ & $\left\{-3,5,-5,3\right\}$   \\
\hline
$\left\{0,1,2,4\right\}$ & $\left\{3,-3,1,-1\right\}+...\; (8)$ & $\left\{-3,8,-6,1\right\}$   \\
\hline
$\left\{0,1,5,6\right\}$ & $\left\{4,0,0,-4\right\}+...\; (26)$ & $\left\{-2,3,-3,2\right\}$   \\
\hline
$\left\{0,1,3,4\right\}$ & $\left\{4,0,0,-4\right\}+...\; (37)$ & $\left\{-1,2,-2,1\right\}$   \\
\hline
$\left\{0,1,2,3\right\}$ & $\left\{4,-1,-1,-2\right\}+...\; (22)$ & $\left\{-1,3,-3,1\right\}$   \\
\hline
$\left\{0,3,5,8\right\}$ & $\left\{4,-2,0,-2\right\}+...\; (18)$ & $\left\{-1,4,-4,1\right\}$   \\
\hline
$\left\{1,3,4,6\right\}$ & $\left\{4,-3,0,-1\right\}+...\; (16)$ & $\left\{-1,5,-5,1\right\}$   \\
\hline
$\left\{0,3,4,7\right\}$ & $\left\{2,-3,2,-1\right\}+...\; (8)$ & $\left\{-1,7,-7,1\right\}$   \\
\hline
$\left\{0,2,3,4\right\}$ & $\left\{1,-1,3,-3\right\}+...\; (8)$ & $\left\{-1,6,-8,3\right\}$   \\
\hline
\end{tabular}
\caption{Table showing the possible charges of the GUT 5-multiplets under the $U(1)$, the possible fluxes along the $U(1)$ and the possible hypercharge fluxes. The charges of the multiplets should be multiplied by a multiple of 5 and shifted by an integer to match the values in F-theory constructions. We considered only charges of values up to 8. In cases when there are many flux combinations the notation is shortened with the number in the brackets giving the number of entries.}
\label{u1charcomp5}
\end{table}

\section{Further models}
\label{sec:extmod}

Having presented the minimal F-theory model, and then a general analysis, in this section we proceed with constructing further models which manifest certain phenomenologically interesting properties. We keep to a rather minimal setting 
of a single U(1) symmetry but allow ourselves the freedom to consider any of the GUT spectra constructed in \cite{Baume:2015wia}. This includes GUT models which are fully realised as F-theory elliptic fibrations but also includes models that can be realised by Higgsing chains from well understood F-theory constructions. We further take the anomaly constraints (\ref{anomNmix}-\ref{anomN}) as the only constraints on hypercharge flux assignments. Note that the most general solution to the anomaly constraints for the models of \cite{Baume:2015wia} was found in \cite{Hahn} which was useful in the model building analysis. 

\subsection{A smaller exotic spectrum}

In this section we construct a model which exhibits the minimal exotic spectrum consistent with gauge coupling unification: a single pair of $5+\overline{5}$ representations. The general analysis of this problem was performed in section \ref{sec:hyp5comp} and we need to find a model to apply those results to. We consider the model labeled $\left\{4,5,4\right\}$ in \cite{Baume:2015wia} which supports the hypercharge flux configuration shown in table \ref{454mod}. The resulting spectrum is shown in table \ref{454mod} where we only show the matter which has some flux restriction to it.
\begin{table}
\centering
\begin{tabular}{|c|c|c|c|c|c|}
\hline
GUT Field & $U(1)$ Charge& $U(1)$ Flux & Hyper Flux & MSSM & Exotics \\
\hline
$10_M$ & -3 & 3 & 0 & $3 \times 10$ & \\
\hline
$5_{E_1}$ & 11 & 1 & -1 & & $\left(3,1\right)_{-\frac13}$\\
\hline
$5_{H_u}$ & 6 & 0 & 2 & $\left(1,2\right)_{\frac12}$ &  $\left(1,2\right)_{\frac12}$\\
\hline
$5_{M}$ & 1 & -3 & 0 & $3 \times \bar{5}$ & \\
\hline
$5_{H_d}$ & -4 & 0 & -2 & $\left(1,2\right)_{-\frac12}$ &  $\left(1,2\right)_{-\frac12}$\\
\hline
$5_{E_2}$ & -9 & -1 & 1 & & $\left(\overline{3},1\right)_{\frac13}$\\
\hline
\end{tabular}
\caption{Table showing the massless spectrum for an F-theory model with the minimal exotic spectrum consistent with gauge coupling unification.}
\label{454mod}
\end{table}
The model is also interesting because the singlets which lift the exotics have charges 10 and 20 and so do not induce a matter-parity violating operator $5_{H_u} \overline{5}_M$ or the proton decay operators $\overline{5}_M \overline{5}_M 10_M$ (due to a remnant $\mathbb{Z}_2$ symmetry) and $\overline{5}_M 10_M 10_M 10_M$. However they do induce a $\mu$-term which ties their vev to the TeV scale.

\subsection{Another combination preserving gauge coupling unification}

In this section we present a model which realises the combination of states (\ref{specexot1}) which preserves gauge coupling unification. We consider the model labeled $\left\{3,4,3\right\}$ in \cite{Baume:2015wia} and the spectrum is presented in table \ref{10sepexmod}.
\begin{table}
\centering
\begin{tabular}{|c|c|c|c|c|c|}
\hline
GUT Field & $U(1)$ Charge& $U(1)$ Flux & Hyper Flux & MSSM & Exotics \\
\hline
$10_M$ & 1 & 4 & -1 & $3 \times 10$ & $(3,2)_{\frac16} +2\times (\bar{3},1)_{-\frac23}$ \\
\hline
$10_{E}$ & 6 & -1 & 1 & &  $\left(\bar{3},2\right)_{-\frac16} +2\times \left(3,1\right)_{\frac23}$  \\
\hline
$5_{H_d}$ & 8 & 0 & -1 & $\left(1,2\right)_{-\frac12}$ & \\
\hline
$5_{H_u}$ & -2 & 0 & 2 & $\left(1,2\right)_{\frac12}$ &  $\left(1,2\right)_{\frac12}$\\
\hline
$5_{M}$ & -7 & -3 & -1 & $3 \times \bar{5}$ &  $\left(1,2\right)_{-\frac12}$\\
\hline
\end{tabular}
\caption{Table showing the massless spectrum for an F-theory model with an exotic spectrum as in (\ref{specexot1}) preserving gauge coupling unification.}
\label{10sepexmod}
\end{table}

\subsection{Separating Standard Model and exotic charges}

In this section we construct a model which is similar to the minimal model of section \ref{sec:minfth} but where we aim to not have the exotic fields have the same quantum numbers as the MSSM matter fields. This could be important for mixing effects and also possibly to generate large Yukawa couplings for the exotics to the singlets as may be required by the 750 GeV excess. In particular such separation can be used to ensure stability of the exotics under decay to the Standard Model fields which could lead to experimental constraints if the exotic masses are light. We consider the model labeled $\left\{4,5,4\right\}$ in \cite{Baume:2015wia} and the fluxes and spectrum are shown in table \ref{10pexmod}.
\begin{table}
\centering
\begin{tabular}{|c|c|c|c|c|c|}
\hline
GUT Field & $U(1)$ Charge& $U(1)$ Flux & Hyper Flux & MSSM & Exotics \\
\hline
$10_{E_1}$ & -8 & 1 & 1 &  & $\left(3,2\right)_{\frac16} + 2 \times\left(1,1\right)_1$\\
\hline
$10_M$ & -3 & 3 & 0 & $3 \times 10$ & \\
\hline
$10_{E_2}$ & 7 & -1 & -1 &  & $\left(\bar{3},2\right)_{-\frac16} + 2 \times\left(1,1\right)_{-1}$\\
\hline
$5_{H_u}$ & 6 & 0 & 1 & $\left(1,2\right)_{\frac12}$ &  \\
\hline
$5_{M}$ & 1 & -3 & 0 & $3 \times \bar{5}$ & \\
\hline
$5_{H_d}$ & -4 & -1 & 0 & $\left(1,2\right)_{-\frac12}$ &  $\left(\bar{3},1\right)_{\frac13}$\\
\hline
$5_{E}$ & -9 & 1 & -1 & & $\left(3,1\right)_{-\frac13}$\\
\hline
\end{tabular}
\caption{Table showing the massless spectrum for an F-theory model where the exotic fields have different quantum numbers to the MSSM matter.}
\label{10pexmod}
\end{table}

\section{Diphoton production rates for F-theory exotics}
\label{sec:prodrates}

Having presented an analysis of the exotics spectra in F-theory models with hypercharge flux, in this section we present an initial more detailed analysis of their potential for explaining the diphoton excess reported in \cite{atlas,cms}. We use the formulae for the gluon-gluon and photon-photon decay rates in \cite{Franceschini:2015kwy}. For a vector-pair of fermions with Yukawa coupling $y_f$ to the scalar component of the singlet $S$, electric charge $L_f$ and mass $M_f$ we have
\bea
\frac{\Gamma\left(S \rightarrow gg \right)}{M} &\simeq& 7.2 \times 10^{-5} \left|\sum_f C_f y_f \frac{M}{2M_f} \right|^2 \;, \nonumber \\
\frac{\Gamma\left(S \rightarrow \gamma\gamma\right)}{M} &\simeq& 5.4 \times 10^{-8} \left|\sum_f d_f L_f^2 y_f \frac{M}{2M_f} \right|^2\;,
\eea
where $M$ is the mass of the singlet, so $750$ GeV, $C_f$ is the Casimir of the $SU(3)$ representation, so $\frac12$ for the fundamental, and $d_f$ is the number of states in the loop. To reproduce the diphoton excess rate we require \cite{Franceschini:2015kwy}
\be
\frac{\Gamma\left(S \rightarrow gg \right)}{M} \sim 10^{-3} - 10^{-6} \;,\;\;  \frac{\Gamma\left(S \rightarrow \gamma\gamma\right)}{M} \sim 10^{-6} \;, \label{prodrate}
\ee
where we assume that the total decay rate for $S$, denoted $\Gamma$, is dominated by $\Gamma_{gg}$ and $\Gamma_{\gamma\gamma}$.
While to reproduce also the width observed by ATLAS we require
\be
\frac{\Gamma}{M} \simeq \frac{\Gamma_{gg}}{M} + \frac{\Gamma_{\gamma\gamma}}{M} \sim 0.06 \;. \label{width}
\ee
We will primarily focus on reproducing the production rates (\ref{prodrate}). Typically we have multiple singlets and multiple exotics which interact in a model dependent way. For simplicity as an initial analysis consider the case of only one singlet coupling to the exotics with a universal coupling. With these approximations the decay rates are determined primarily by the exotics spectrum which is sufficient as an approximate initial analysis. The decay rates for the exotics spectra which arise in the models are shown in table \ref{diphoton}. In the final column we show the value of the universal Yukawa coupling necessary to match a production rate of $\Gamma_{\gamma\gamma}/M=10^{-6}$ if the universal exotics mass is $400$ GeV. These remain perturbative for all the possibilities apart from a single pair of 5-multiplets. Note that these can be relaxed by taking a smaller production rate or smaller exotics mass. In more detailed analyses of models we may have multiple singlet states coupling to subsectors of the exotics which will typically lead to slightly reduced production rates in a model dependent way. However the sectors of exotics coming from the 10 multiples will couple to just one singlet and so in this sense form a building block of the more general production rates and can be used as approximate lower bounds on these. We therefore also include the calculation for these two subsectors in table \ref{diphoton}.
\def\arraystretch{1.5}
\begin{table}
\centering
\begin{tabular}{|C{5.8cm}|C{2.3cm}|C{2.3cm}|C{1.7cm}|C{1.7cm}|}
\hline
Exotics & $X_{gg}$ & $X_{\gamma\gamma}$ & $\left.y_f\right|_{400 \mathrm{GeV}}$ & $\left.\frac{\Gamma}{M}\right|_{400 \mathrm{GeV}}$\\
\hline
$N \times 5$ & $1.8 \times 10^{-5} N^2 $ & $9.6 \times 10^{-8}N^2$ & $1.7/N$ & $1.8 \times 10^{-4}$\\
\hline
$\left[ (3,2)_{+\frac16} + (\bar{3},1)_{+\frac13} + 2(1,1)_{+1} \right]$ & $1.6 \times 10^{-4}$ & $8.6 \times 10^{-7}$ & $0.58$ &$1.9 \times 10^{-4}$\\
\hline
$\left[ (3,2)_{+\frac16} + (\bar{3},1)_{+\frac13} + 2(1,1)_{+1} +5\right]$  & $2.9 \times 10^{-4}$ & $1.5 \times 10^{-6}$ & $0.43$ & $1.9 \times 10^{-4}$ \\
\hline
$\left[ (3,2)_{+\frac16} + 2  (3,1)_{+\frac23} + (1,2)_{+\frac12} \right]$  & $2.9 \times 10^{-4}$ & $1.5 \times 10^{-6}$ & $0.43$ & $1.9 \times 10^{-4}$\\
\hline
$\left[ (3,2)_{+\frac16} + 2(1,1)_{+1} \right]$  & $7.2 \times 10^{-5}$ & $7.3 \times 10^{-7}$ & $0.62$ & $9.7 \times 10^{-5}$\\
\hline
$\left[ (3,2)_{+\frac16} + 2  (3,1)_{+\frac23}  \right]$  & $2.9 \times 10^{-4}$ & $1.0 \times 10^{-6}$ & $0.53$ & $2.9 \times 10^{-4}$\\
\hline
\end{tabular}
\caption{Table showing the decay rates for different exotics spectra. Note that the spectra are completed with the conjugate states. The decay rates are of the form $\frac{\Gamma_{gg}}{M} \simeq X_{gg} \left|y_f\frac{M}{M_f}\right|^2$. The final two columns show the value of the universal Yukawa coupling necessary for matching the production rate of $\Gamma_{\gamma\gamma}/M=10^{-6}$ for $M_f=400$GeV and the total width for such a coupling.}
\label{diphoton}
\end{table}
\def\arraystretch{1.0}
If we assume that the total width is dominated by the gluons and photons decay channels then it appears difficult to reproduce a large width (\ref{width}). In table \ref{diphoton} we show the width for the value of the coupling fixed by requiring the correct production rate for $M_f=400$ GeV. It is a couple of orders of magnitude too small. It is a fairly conservative value and it is likely that it is possible to increase the magnitude but probably not to $0.06$. One way to explain this is by allowing for a small split in the singlets masses, of order $\sim 40$ GeV, which could explain the large width observed by ATLAS as multiple nearly degenerate resonances \cite{Franceschini:2015kwy}. Another possibility is that the singlet decays primarily to other states $\Gamma  = \Gamma_{gg}+\Gamma_{\gamma\gamma}+\Gamma_{X}$ with $\Gamma_{X} \sim 0.06$. In the F-theory constructions this is a natural possibility due to the presence of further charged singlets which do not couple to the exotics but do have Yukawa couplings with $S$. Note that there may be constraints on this possibility from monojet signatures \cite{Franceschini:2015kwy}. We leave a more detailed analysis of the experimental signitures and implications of such couplings for future work.

\section{Summary}
\label{sec:summary}

In this note we studied the spectrum of vector-like exotics, and singlets which couple to them, which arise inevitably for a large class of F-theory GUTs with a U(1) symmetry under which the MSSM Higgs fields are not vector-like and where hypercharge flux induces doublet-triplet splitting. We showed that already the minimal implementation of such a U(1) symmetry predicts a spectrum of exotics beyond the MSSM which couple to singlets and preserve gauge coupling unification. The exotics are vector-like with respect to the SM gauge symmetries but not with respect to the U(1) symmetry. The breaking of the U(1) symmetry and subsequently the exotics mass can be naturally tied to the TeV scale through the $\mu$-term, proton decay and matter-parity violation constraints. This presents an ultraviolet motivation for the presence of a spectrum of fields that may explain the possible excess of diphoton events observed at the LHC at around 750 GeV. 

We showed that constraints on the spectrum of exotics induced through hypercharge flux are sufficiently strong that a general analysis of the possible ways that the spectrum can arise can be performed. The results of this analysis presented a distinctive and predictive spectrum of fields and relations between the U(1) charges of the exotics. The U(1) charges in turn predict the structure of the coupling of the vector-like charged fields to the singlet fields and can be tested experimentally. 

Using the general results we then presented a number of non-minimal F-theory GUT models with a U(1) symmetry that exhibit certain phenomenological features. For example realising the different types of exotics spectra which can maintain gauge coupling unification and suppressing mixing effects between the exotics and MSSM fields. This was a first run at more sophisticated model building with vector-like exotics at the TeV scale in F-theory GUTs. Future work in this direction, for example by utilising more than one U(1) symmetry, will likely yield ever more phenomenologically realistic models. 

The analysis presented in this work can be motivated independently of the possible excess observed at the LHC. Nonetheless the excess presented a key motivation for this work. We presented an initial analysis of whether the spectrum of exotics present can explain the data of the LHC. We find that reproducing the diphoton production rates can be done with natural values of the model parameters. The hints of a large decay width seen by ATLAS may be difficult to accommodate if they continue to hold. Though the feature of many of the models of having multiple singlets may be helpful with this. We leave a more detailed phenomenological study for the future and in particular for after the fate and precise details of the excess are better understood with more data. 

{\bf Acknowledgments}
The work of EP is supported by the Heidelberg Graduate School for Fundamental Physics. I thank Emilian Dudas for useful discussions.

\appendix
\section{Classifying solutions to the hypercharge constraints}
\label{sec:app}

In this appendix we outline the methodology used to classify the possible fluxes and U(1) charges that can lead to the spectrum of exotics (\ref{specexot2}). We split the possibilities into the case where one of the hypercharge fluxes over the 5-matter curves vanishes and when they are all non-vanishing. We start with the latter. In that case we can solve (\ref{anomNmix}-\ref{anomN}) as
\bea
& &N_{10}^1 = -N_{10}^2 = -\frac{A B N_5^3}{C \left( 3C -A - B - 2D\right)} \;, \nonumber \\
& &N_5^1 = \frac{B \left(B - 3C +2D\right)N_5^3}{\left(A-B\right)\left(A+B-3C+2D\right)}\;,\;N_5^2 = -\frac{A \left(A - 3C +2D\right)N_5^3}{\left(A-B\right)\left(A+B-3C+2D\right)} \;. \label{solN}
\eea
Where we define
\be
A = Q_5^1-Q_5^3 \;,\; B = Q_5^2-Q_5^3 \;,\; C = Q_{10}^1-Q_{10}^2\;,\; D = Q_5^3 - 3 Q_{10}^2 \;. 
\ee
Note that since $N_5^3$ is non-vanishing, and since $\left(B - 3C +2D\right)$ and $\left(A+B-3C+2D\right)$ can not both simultaneously vanish, this solution is valid as long as $\left(A+B-3C+2D\right) \neq 0$. We can now scan over the possible charges and classify the possible integer solutions to these equations. It is more convenient to scan over the values of $A,B,C$ and $D$ because the solutions are independent under an overall rescaling of them which means we can divide them by 5 so that they take values in the integers. Further we are free with full generality to assume $A>B>0$ and $C>0$. Their bounding top values are taken from the assumptions over the maximum separation of the charges discussed above, which implies $1 \leq A \leq 7$ and $1 \leq C \leq 4$. To determine the range for $D$ we note that after rescaling $\left|D\right| = \frac15\left|Q_5^3-3 Q_{10}^1 \right| \leq \frac15\left(\left|Q_5^3\right|+3 \left|Q_{10}^1\right| \right) \leq 27$, where we assumed that at least one of the charges of each type of state is no further from zero than 5.\footnote{Possibly after an appropriate overall shift $Q_{10}^i \rightarrow Q_{10}^i + a$ and $Q_{5}^i \rightarrow Q_{5}^i + 3a$ under which the system is invariant.} Finally we need to scan over $N_5^3$ which we can take to be positive with generality and also must satisfy $N_5^3 \leq 4$ to get the spectrum (\ref{specexot2}). We can now perform a computer scan over the possible integer values of $A$, $B$, $C$ and $D$ and pick out the values which allow for integer hypercharge fluxes. This gives the possible values of the U(1) charges and hypercharge fluxes. With these we then scan over possible values of U(1) fluxes that can lead to the spectrum (\ref{specexot2}). This then leads to the results presented in table \ref{u1char10}.

Now consider the case where one of the hypercharge fluxes over the 5-curves vanishes. Without loss of generality we can choose this to be $N_5^3=0$. Now the constraints (\ref{anomNmix}-\ref{anomN}) are four linear equations in only four variables and so imply vanishing of all the fluxes unless a constraint is imposed on the charges. In the notation above this is $A+B-3C+2D = 0$, which picks out the set of solutions missed out in the analysis above. With this constraint we have the solution
\be
N_{10}^2 = -N_{10}^1 \;,\; N_{5}^1 = -N_{5}^2 = \frac{C N_{10}^1}{B-A} \;.
\ee 
We can now perform a similar scan as before but only over $A$, $B$ and $C$ which leads to a much more constrained set of solutions (note that without loss of generality we can choose $N_{10}^1=1$). The results are presented in table \ref{u1char10}.


\end{document}